\documentclass[preprint,12pt]{elsarticle}

\usepackage{graphics}
\usepackage{amsfonts}
\usepackage{bm}
\usepackage{amsmath}
\usepackage{amssymb}
\usepackage{color}
\usepackage[all]{xy}
\usepackage{tikz}
\usepackage{tikz-cd}

\def\be{\begin{equation}}
\def\ee{\end{equation}}
\def\bea{\begin{eqnarray}}
\def\eea{\end{eqnarray}}

\journal{Journal of Geometry and Physics}

\begin{document}

\begin{frontmatter}

\title{On the physical significance of geometrothermodynamic metrics}

\author{V. Pineda$^{1}$, H. Quevedo$^{1,2,3}$, M. N. Quevedo$^{4}$, Alberto S\'anchez$^{5}$, 
and Edgar Vald\'es$^{1}$}
\ead{viridiana.pineda@correo.nucleares.unam.mx, quevedo@nucleares.unam.mx, maria.quevedo@unimilitar.edu.co, asanchez@nucleares.unam.mx, 
edgar.valdes@isaptechnologies.com}

\address{$^1$Instituto de Ciencias Nucleares, Universidad Nacional Aut\'onoma de M\'exico,
 AP 70543, M\'exico, DF 04510, Mexico\\
$^2$Dipartimento di Fisica and ICRANet, Universit\`a di Roma ``La Sapienza",  I-00185 Roma, Italy\\
$^3$Department of Theoretical and Nuclear Physics, Kazakh National University, Almaty 050040, Kazakhstan\\
$^4$Departamento de Matem\'aticas, 
 Facultad de Ciencias B\'asicas, 
Universidad Militar Nueva Granada, Cra 11 No. 101-80, 
Bogot\'a D.E., Colombia \\
$^5$Departamento de Posgrado, CIIDET,  AP752, 
Quer\'etaro, QRO 76000, Mexico
}

%

\begin{abstract}
In geometrothermodynamics (GTD), to study the geometric properties of the equilibrium space  three thermodynamic metrics have been proposed so  far. 
These metrics are obtained by using the condition of Legendre invariance and can be computed explicitly once a thermodynamic potential is 
specified as fundamental equation.  We use the remaining diffeomorphism invariance in the phase and equilibrium spaces to show that 
the  components of the GTD-metrics can be interpreted as the second moment of the fluctuation of a new thermodynamic potential. 
This result establishes a direct connection between GTD and fluctuation theory. In this way, the diffeomorphism invariance of GTD allows us to introduce 
new thermodynamic coordinates and new thermodynamic potentials, which are not related by means of  Legendre transformations to the fundamental thermodynamic potentials.
\end{abstract}


\end{frontmatter}

\section{Introduction}
\label{sec:int}

Classical equilibrium thermodynamics is based upon a set of empirical observations that became the laws of thermodynamics. 
This empirical background is probably the cause of the tremendous success of thermodynamics in all possible configurations that can be considered as thermodynamic systems. Although usually a thermodynamic system is specified through its equations of state, an alternative method consists in specifying the fundamental equation from which all the equations of state can be derived \cite{callen}. In turn, a fundamental equation is determined by means of a thermodynamic potential $\Phi$ which is usually understood as a function of the extensive variables $E^a$, 
$a=1,...,n$, where $n$ is the number of thermodynamic degrees of freedom of the system. Furthermore, it is well known that from a given thermodynamic potential $\Phi$, it is possible to obtain a set of new potentials $\tilde \Phi$ by means of Legendre transformations. 
An important result of classical thermodynamics is that the properties of a system do not depend on the choice of potential. From a 
theoretical point of view this implies that classical thermodynamics is invariant with respect to Legendre transformations, i.e., it 
preserves Legendre invariance. 
 
On the other hand, the utilization of differential geometry in classical thermodynamics has shown to be of special importance as an alternative representation and as an additional tool to investigate stability and criticality properties of thermodynamic systems 
\cite{amari85,wei09,rup14}. The main idea consists in representing all possible equilibrium states of a system as points of an abstract space that is called equilibrium space. Furthermore, the main goal is to find a relationship between the geometric properties of the equilibrium space and the thermodynamic properties of the system.  In this regard, the approach based upon the introduction of specific metrics into the equilibrium space is known as thermodynamic geometry. In particular, one can use the second derivatives of a thermodynamic potential to construct a Hessian matrix which determines a Hessian metric, if its determinant is non-zero. In this way, the equilibrium space becomes a Riemannian manifold with well-defined connection and curvature. One then tries to associate  connection and geodesics with quasi-static processes, curvature with thermodynamic interaction and curvature singularities with phase transitions. The interesting issue about using Hessian metrics in the equilibrium space is that they have a very remarkable physical significance. In fact, if we consider small fluctuations of the thermodynamic potential, which should be an extremal at each equilibrium point, the second moment of the fluctuation turns out to be directly related to the components of the corresponding Hessian metric. This is, of course, a very important result of thermodynamic geometry.

A different approach is represented by the formalism of geometrothermodynamics (GTD) whose main feature is that it takes into account the
Legendre invariance of classical thermodynamics, a property which is not considered in thermodynamic geometry. To this end, GTD uses the thermodynamic phase space as an auxiliary manifold. However, the GTD-metrics are not Hessian and, therefore, their physical interpretation is not clear. The main goal of the present work, is to show that in GTD it is possible to use the freedom in choosing the coordinates of the phase and equilibrium spaces in order to introduce a new thermodynamic potential and new thermodynamic coordinates such that the metric of the equilibrium space becomes Hessian. This proves that the components of the GTD-metrics can be also interpreted as the second moment of fluctuation of a thermodynamic potential. 

This work is organized as follows. In Sec. \ref{sec:geomet}, we present a brief review of the formalisms of thermodynamic geometry and GTD, including the general form of the metrics that are used in both formalisms.  In Sec. \ref{sec:diff}, we use the diffeomorphism invariance that is left in the phase and equilibrium spaces to obtain the conditions for the GTD-metrics to become Hessian. First, we consider diffeomorphisms that act on the phase space, and introduce the concept of deformed contactomorphisms in order to integrate the differential equations that determine the diffeomorphism. We show that there exists a particular diffeomorphism that transforms the partially and totally Legendre invariant metric of GTD into a Hessian metric at the level of the equilibrium space. The Hessian potential turns out to be a new coordinate introduced by the particular diffeomorphism of the phase space. We then consider the GTD-metrics that are invariant with respect to total Legendre transformations only. We show in Sec. \ref{sec:hess} that in this case it is necessary to consider the diffeomorphism invariance of the equilibrium space in order to introduce new coordinates that transform the GTD-metrics into Hessian metrics. It turns out that the existence of the Hessian potential in this case depends on the dimension $n$ of the equilibrium space. If $n=2$, there always exists a Hessian potential, but for $n\geq 3$ a Hessian potential exists only if the Riemann curvature of the equilibrium space satisfies certain algebraic conditions related to the Pontryagin characteristic class of the manifold. 
It is then necessary to consider each particular thermodynamic system separately in order to establish the existence of a Hessian potential.
In Sec. \ref{sec:pot}, we propose a new classification of thermodynamic potentials according to their thermodynamic and geometric properties. We classify thermodynamic potentials into fundamental, Legendre and diffeomorphic potentials. Finally, in Sec. \ref{sec:con} 
 we discuss our results.

\section{Hessian metrics and GTD-metrics }
\label{sec:geomet}

Consider an analytic function $H(E^a)$, $a=1,2,...,n$. In an $n-$dimensional manifold, it is possible to introduce a Hessian metric as
\be 
g^H = \frac{\partial^2 H}{\partial E^a\partial E^b} dE^a dE^b \ ,
\label{hess}
\ee  
if the condition det$(g^{^H}_{ab})\neq 0$ is satisfied. If we now consider the set $\{E^a\}$ as the coordinates of a differential manifold ${\cal E}$ and the function $H(E^a)$ as a thermodynamic fundamental equation, then ${\cal E}$ can be interpreted as an equilibrium manifold  with the Hessian metric (\ref{hess}) determining all the geometric properties of ${\cal E}$.
The function $H(E^a)$ is also known as the Hessian potential.
 This is exactly the procedure that is followed in thermodynamic geometry to endow the equilibrium space with a metric. 

The physical meaning of the components of Hessian metrics justifies its utilization and leads to a large number of 
physical applications. Indeed, since a set of values $E^a$ represents a state of equilibrium in ${\cal E}$, where from a physical point of view fluctuations are always present, let us denote by $dE^a$ the infinitesimal deviations of the variables $E^a$ from the equilibrium state.
To take into account the fluctuations of $H(E^a)$ around an equilibrium state, one basically analyzes the behavior of the Taylor expansion 
\cite{ll77} 
\be
H(E^a+dE^a) = H(E^a) + \frac{\partial H}{\partial E^a} d E^a + \frac{1}{2} \frac{\partial^2 H}{\partial E^a\partial E^b } dE^a d E ^b + \cdots 
\label{tay}
\ee
At equilibrium, the thermodynamic potential reaches an extremal value which can be a maximum or a minimum, depending on the type of thermodynamic potential. Then, the first-order derivatives vanish at equilibrium and the second moment of the fluctuation 
corresponds basically to the components of the Hessian metric. This can be considered as the mathematical definition of a thermodynamic potential in the framework of fluctuation theory; we will use this definition for the analysis of the new thermodynamic potentials that will be introduced below. 

From a physical point of view, however, the interpretation of $H(E^a)$ as a thermodynamic potential requires the consideration of the properties of the underlying thermodynamic system. In fact, for analyzing the physical processes which can occur in a particular   thermodynamic system it is necessary to consider it as a part of a larger system which includes the reservoir. At equilibrium, the properties of both the system and the reservoir must be considered and if the additivity condition is satisfied, the potential $H(E^a)$ splits into two pieces, each possibly with different properties. Then, different arguments must be invoked to show that the second term of the Taylor expansion (\ref{tay}) vanishes at equilibrium. In turn, the arguments depend on the particular physical significance of the potential \cite{ll77}. For instance, if we choose $H$ as the entropy $S$, then, it must be identified as the entropy of the 
entire Universe. Moreover, since it reaches a maximum at equilibrium, i.e., $\partial S/\partial E^a=0$,  the second fluctuation  moment 
given in Eq.(\ref{tay}) corresponds basically to the Hessian matrix of the entropy which is known as the Ruppeiner metric \cite{rup14}, and is
also interpreted as the stability matrix in fluctuation theory. This important result provides Ruppeiner metric  with a clear physical significance, and permits to find the connection with information geometry. Furthermore, if we choose $H$ as the internal energy $U$, 
a similar argumentation can be used to show that the second term in the Taylor expansion (\ref{tay}) vanishes, and the second moment of the fluctuation $\frac{\partial^2 U}{\partial E^a\partial E^b }$ 
determines the components of the Weinhold metric \cite{wei09}, which is related to the Ruppeiner metric by means of a conformal factor. 
The above description shows the important physical significance of Hessian metrics in thermodynamic geometry. 

In GTD, the procedure for  introducing Riemannian metrics into the equilibrium space is different. GTD aims to incorporate 
into a geometric description all the mathematical and physical properties of classical thermodynamics. Therefore, it is necessary
to introduce additional structures that make it possible to understand the laws of thermodynamics from a geometric point of view.
In the following description, we will use the Legendre invariance property of thermodynamics in order to derive the metrics that are used in GTD to describe physical systems.

In classical thermodynamics, to describe an arbitrary system with $n$ degrees of freedom, 
it is necessary to consider $n$ extensive variables $E^a$,  $n$ intensive variables $I^a$, and one thermodynamic potential $\Phi$. 
For a particular system, the corresponding fundamental equation must be specified, which is usually represented by a function that relates the potential with the extensive variables, i.e., $\Phi=\Phi(E^a)$. For the sake of generality, we will demand that $\Phi(E^a)\in C^\infty$ is an analytical function. The first law of thermodynamics implies then that (summation over repeated indices)
\be
d\Phi = I_a d E^a \ , \quad I_a = \frac{\partial \Phi}{\partial E^a} \ .
\label{flaw}
\ee
In this way, the intensive variables become explicitly functions of the extensive variables, i.e., $I_a = I_a (E^b)$. 

A Legendre transformation is usually represented as a change of potential that includes its first-order derivatives. Consider, for instance, the simple case of a function of only one variable $\Phi=\Phi(E^1)$. A Legendre transformation of $\Phi$ implies the introduction of a new potential 
$\tilde \Phi$ by means of 
\be
\Phi \rightarrow \tilde \Phi = \Phi - \frac{\partial \Phi}{\partial E^1}  E^1  \ .
\label{leg0}
\ee
Then, from the first law of thermodynamics we obtain $d\tilde \Phi = - E^1 dI_1$, implying that 
\be 
\tilde \Phi=\tilde \Phi(I_1), \quad {\rm  and} \quad E^1 = - \frac{\partial\tilde\Phi} {\partial I_1} \ .
\ee
This means that a Legendre transformation affects the functional dependence of the potential and interchanges the role of extensive and intensive variables. 

Our goal is to represent Legendre transformations in a pure differential geometric fashion, for instance, as a coordinate transformations.
Clearly, the Legendre transformation (\ref{leg0}) cannot be interpreted as a change of coordinates because it includes the derivatives of one of the coordinates. It is therefore necessary to introduce an auxiliary space in which all the coordinates are independent. Let us consider the ($2n+1)-$dimensional space ${\cal T}$ with coordinates $Z^A=\{\Phi, E^a, I_a\}$, where $A=0,1,...,2n$. Then, the coordinate 
transformation \cite{arnold}
\be
\{Z^A\}\longrightarrow \{\widetilde{Z}^A\}=\{\tilde \Phi, \tilde E
^a, \tilde I ^ a\}\ , 
\ee
 \be
 \Phi = \tilde \Phi - \tilde E ^k \tilde I_k  \ ,\quad
 E^i = - \tilde I ^ {i} \ , \ \
E^j = \tilde E ^j,\quad
 I^{i} = \tilde E ^ i , \ \
 I^j = \tilde I ^j \ ,
 \label{leg}
\ee 
where $i\cup j$ is any disjoint decomposition of the set of
indices $\{1,...,n\}$, $k,l= 1,...,i$. For
$i=\emptyset$, we obtain the identity transformation, and for $i=\{1,...,n\}$ a total Legendre
transformation, i.e., 
\be
 \Phi = \tilde \Phi -  \tilde E ^a \tilde I _b \ ,\quad
 E^a = - \tilde I ^ {a}, \ \
 I^{a} = \tilde E ^ a \ .
 \label{legtotal}
\ee 
Here we use the notation $I^a = \delta^{ab} I_b$ with $\delta_{ab} = {\rm diag}(1,...,1)$.   It is then easy to see that if we introduce 
the dependence $\Phi=\Phi(E^a)$, the above coordinate transformation reduces to the Legendre transformation which includes derivatives 
of $\Phi$. From now on we refer to Eq.(\ref{leg}) as Legendre transformations. 

Let us now suppose that ${\cal T}$ is a smooth manifold. Then, by virtue of Darboux's theorem \cite{stern99}, the space ${\cal T}$ with coordinates $Z^A=\{\Phi, E^a, I_a\}$ can be endowed with a unique 
(up to a conformal function) differential 1-form $\Theta = d\Phi - I_a dE^a$ which satisfies the condition 
$\Theta \wedge (d\Theta)^n \neq 0$, and is called  contact 1-form.  The pair $({\cal T}, \Theta)$ is known as contact manifold. 
The interesting point about the canonical contact 1-form is that it is invariant with respect to Legendre transformations, i.e., 
the new 1-form $\tilde \Theta$, which is obtained from $\Theta$ by applying a Legendre transformation (\ref{leg}), can be written
as $\tilde \Theta = d\tilde \Phi - \tilde I _ a d\tilde E^a$, implying that the functional dependence of $\Theta$ remains unchanged.
This means that in general a contact manifold is Legendre invariant.

We now introduce a Riemannian metric structure $G$ into the contact manifold. The triad $({\cal T}, \Theta, G)$ 
forms an odd-dimensional Riemannian contact manifold which is the basis for the construction of GTD. Indeed, if 
we demand that the metric $G$ be Legendre invariant,  i.e., the functional dependence of $G= G_{AB} d Z^A dZ^B$ should remain 
invariant under the action of Legendre transformations. The triad $({\cal T}, \Theta, G)$ constitutes the thermodynamic phase space 
of GTD \cite{quev07}. This completes the construction of a Legendre invariant formalism at the level of the phase space which, as we have shown, is an auxiliary manifold we introduced in order for Legendre transformations to be considered as coordinate transformations. 
The physical space where the thermodynamic systems can be investigated is represented by the equilibrium space ${\cal E}$ which is defined
as a submanifold of ${\cal T} $ by means of the embedding smooth map $\varphi: {\cal E} \rightarrow {\cal T}$ such that the pullback 
$\varphi^*$ satisfies the condition $\varphi^*(\Theta)=0$. Furthermore, if we choose the set $\{E^a\}$ as the coordinates of the submanifold ${\cal E}$, then the embedding map reads $\varphi: \{ E^a\}\mapsto \{\Phi(E^a), E^a, I^a(E^a)\}$, and the  condition  
$\varphi^*(\Theta)=0$  implies the first law of thermodynamics (\ref{flaw}). Notice that the specification of the map $\varphi$ implies that the fundamental equation must be given explicitly. 
In addition, the embedding map $\varphi$ induces a canonical 
metric $g$  on ${\cal E}$ by means of $g=\varphi^*(G)$. One of the aims of GTD is to describe the physical properties of a thermodynamic system by using the geometric properties of the corresponding Riemannian equilibrium manifold  $({\cal E}, g)$. 

The above construction of the GTD equilibrium manifold shows that it is completely determined once the metric $G$ and the fundamental equation $\Phi(E^a)$ are given. It is therefore necessary to find the explicit form of $G$ which is invariant under Legendre transformations. 
In GTD, we impose an additional physical condition on $G$. In the equilibrium manifold, the metric $g$ is expected to represent the properties of the system; in particular, we want to identify the curvature of ${\cal E}$ as a measure of thermodynamic interaction. This means that ${\cal E}$ must be flat for a system without thermodynamic interaction, i.e., for the ideal gas. Clearly, because of the relation  $g=\varphi^*(G)$, this demand implies a condition on the form of $G$. To carry out this procedure, the problem appears that Legendre transformations are discrete and therefore not all of them can be generated from infinitesimal generators. Moreover,
if we limit ourselves to infinitesimal Legendre transformations, the corresponding metric $G^{inf}$  for the phase space does not lead to 
a flat metric $g^{inf}=\varphi^*(G^{inf})$ for the ideal gas \cite{gl14,qqs16}. Consequently, it is necessary to apply the general form of the transformation 
(\ref{leg}) to an arbitrary metric $G=G_{AB}dZ^A dZ^B$ and demand the functional dependence invariance. The resulting set of algebraic equations \cite{quev07} can be solved for the metric components $G_{AB}$ and then the flatness  condition for the corresponding 
metric $g_{ab}$ can be imposed. As a result, we obtain two different classes of solutions  
\be
 G^{^{I/II}} = (d\Phi - I_a d E^a)^2 + (\xi_{ab} E^a I^b) (\chi_{cd} dE^c dI^d) \ ,
\label{gupap}
\ee
which are invariant under total Legendre transformations and a third class (summation over all repeated indices)
\be	
	\label{GIIIL}
	G^{^{III}}  =(d\Phi - I_a d E^a)^2  +  \left(E_a I_a \right)^{2k+1}  d E^a   d I^a \ , \quad k\in \mathbb{Z}\ ,
\ee
which is invariant under partial Legendre transformations. Here $\xi_{ab}$ and $\chi_{ab}$ are diagonal constant $(n\times n)$-matrices. 
If we choose $\chi_{ab} = \delta_{ab}= {\rm diag}(1,\cdots,1)$, the resulting metric $G^{^I}$ can be used to investigate systems with at least one first-order phase transition. Alternatively, for 
$\chi_{ab} = \eta_{ab}= {\rm diag}(-1,\cdots,1)$, we obtain a metric $G^{^{II}}$ that correctly describes systems with second-order phase transitions. 

To consider the three classes of Legendre 
invariant metrics in a compact form, we introduce the notation 
\be
G = \Theta^2 + h_{ab} dE^a dI ^b \ ,
\label{gupg}
\ee
where the components $h_{ab}$ are in general functions of $E^a$ and $I^a$. Then, the general form of the metric for the equilibrium manifold 
can be written as 
\be
g=\varphi^*(G) = h_{ab} \delta^{bc} \frac{\partial^2 \Phi}{\partial E^c \partial E^d} dE^a dE^d \ ,
\label{gdown}
\ee
which is not a Hessian metric. 

In contrast to the above construction of GTD, in the alternative approach of thermodynamic geometry, the equilibrium manifold is equipped with a Hessian metric
\be
g^H = \frac{\partial^2 \Phi}{\partial E^a \partial E^b} dE^a dE^b \ ,
\ee
which is not invariant with respect to Legendre transformations. Nevertheless, the components of a Hessian metric have a very clear physical interpretation in fluctuation theory, as described above. The metrics used in GTD (\ref{gdown}) are not Hessian due to the presence of the conformal factor determined by the components of $h_{ab}$, which on ${\cal E}$ are in general functions of the potential 
$\Phi$ and its derivatives.


\section{Diffeomorphism invariance}
\label{sec:diff}

The GTD-metrics of the phase and equilibrium manifolds were obtained under the condition that they preserve Legendre invariance. 
Their functional dependence is fixed, but we can still perform coordinate transformations without affecting their geometric properties.
For instance, in the equilibrium manifold we can apply a diffeomorphism $\chi : {\cal E}\rightarrow {\cal E}$, or in coordinates
$\chi : \{E^a\} \mapsto \{E^{\bar a}\}$, which preserves the properties of ${\cal E}$. In this work, we will focus only on  diffeomorphism which are equivalent to coordinate transformations.

It is clear that in general it is not possible to transform an arbitrary metric like (\ref{gdown}) into a Hessian metric by using only the diffeomorphism $\chi$. However, the phase manifold offers more possibilities because it is also invariant with respect to coordinate transformations of the form $\chi_{_{\cal T}}: \{Z^A\}\mapsto \{Z^{\bar A}\}$. For concreteness, let us consider 
$\{Z^{\bar A}\} = \{F, X^a, Y^a\}$, where in general $F=F(\Phi, E^b, I^b)$, $X^a= X^a (\Phi, E^b, I^b)$, and $Y^a=Y^a(\Phi, E^b, I^b)$.  
In turn, the diffeomorphism $\chi_{_{\cal T}}$ induces a diffeomorphism $\chi_{_{\cal E}}$ on ${\cal E}$, according to the  
diagram represented in Fig. 1.

\begin{center}
\begin{tikzpicture}[description/.style={fill=white,inner sep=4pt}]
\matrix (m) [matrix of math nodes, row sep=3em,
column sep=2.5em, text height=1.5ex, text depth=0.25ex]
{ ({\cal T}, Z^A) & & ({\cal T}, Z^{\bar A}) \\
 ({\cal E}, E^a)  &  &   ({\cal E},X ^a) \\ };
\path[->,font=\scriptsize]
(m-1-1) edge node[auto] {$ \chi_{_{\cal T}} $}  (m-1-3) 
(m-2-1) edge node[left] {$ \varphi_{_E}   $} (m-1-1)
(m-2-3) edge node[right] {$ \bar \varphi _{_X} $} (m-1-3)
(m-2-1) edge node[below] {$ \chi_{_{\cal E}}  $} (m-2-3)
;
\end{tikzpicture}
\end{center}
Fig. 1: Diagram of maps relating the phase and equilibrium manifolds.

\

\

Our goal is to apply a general diffeomorphism $\chi_{_{\cal T}}$ on ${\cal T}$ such that the resulting metric can be written as
\bea
\chi_{_{\cal T}}(G) = \bar G & = & \chi_{_{\cal T}}[(d\Phi-I_adE^a)^2 + h_{ab} dE^a dI^b] \nonumber \\
& = &\bar G _1 + \delta_{ab} dX^a dY^b \ ,
\label{cond1}
\eea
where the first component $\bar G _ 1$ satisfies the condition 
\be
\bar \varphi^*_{_X} (\bar G _ 1) = 0\ .
\label{cond2}
\ee
The reason for demanding this 
particular form of the metric is as follows. First, let us note that, according to Darboux's theorem,
 the introduction of the coordinates $\{Z^{\bar A}\} = \{F, X^a, Y^a\}$ implies that there exists a canonical contact form 
$\Theta^F = f(dF - Y_a dX^a)$, where $f$ is a non-zero function  $f: {\cal T} \rightarrow \mathbb{R}$. Then, we can introduce 
a smooth map $\bar \varphi _{_ X }: {\cal E} \rightarrow {\cal T}$ such that $\bar \varphi ^*_ {_ X }(\Theta^F) = 0$, implying 
that 
\be
dF = Y_a dX^a=  \frac{\partial F}{\partial X^a} dX^a 
\ee 
on ${\cal E}$. Consequently,  taking into account the condition 
$\bar \varphi_{_X}^* (\bar G _ 1) = 0$, it can be seen that 
the metric induced by the pullback $\bar \varphi^*_{_X}$ on ${\cal E}$ reads
\be
\bar g = \bar \varphi^*_{_X} (\bar G ) = \bar \varphi^*_{_X} ( \delta_{ab} dX^a dY^b ) =
\frac{\partial^2 F}{\partial X^a \partial X^b} dX^a dX^b\ ,
\ee 
i.e., it is a Hessian metric. This would imply that in principle we can associate the components of the GTD-metrics with the second fluctuation moment
of $F$, if it can be considered as a thermodynamic potential.

We notice that the conditions (\ref{cond1}) and (\ref{cond2}) are sufficient for the induced metric $\bar g$ to become Hessian. In the following sections, we will analyze this condition in the particular case of the GTD-metrics.

\subsection{Deformed contactomorphisms}
\label{sec:def}

To perform the above procedure for determining the explicitly form of the diffeomorphism $\chi_{_{\cal T}}$, it is necessary to
carry out lengthy computations involving partial differential equations. The method rapidly becomes cumbersome and difficult to be carried out. To facilitate the analysis of the problem, we make some simplifying assumptions. When applying the diffeomorphism $\chi_{_{\cal T}}$ on the general form of the metric $G$ given in Eq.(\ref{gupg}), let us assume that the following conditions are satisfied
\bea
\chi_{_{\cal T}}(d\Phi - I_a dE^a) & = & f(dF - Y_a dX^a)\ , \label{cont1}\\  
\chi_{_{\cal T}}(h_{ab}dE^a dI^b) &  =  & \delta_{ab} dX^a dX^b \ .\label{cont2}
\eea
These assumptions satisfy identically the general conditions (\ref{cond1}) and (\ref{cond2}). If the above assumptions are satisfied, the map
$\chi_{_{\cal T}}$ becomes a contactomorphism, i.e., a diffeomorphism that preserves the contact structure. The first assumption 
(\ref{cont1}) leads to  a system of $2n+1$ partial differential equations
\be 
\frac{\partial F}{\partial \Phi} =\frac{1}{f} + Y_a \frac{\partial X^a}{\partial \Phi} \ ,\quad
\frac{\partial F}{\partial E^a} = -\frac{I_a}{f} + Y_b \frac{\partial X^b}{\partial E^a} \ ,\nonumber
\ee
\be
 \frac{\partial F}{\partial I^a} = Y_b \frac{\partial X^b}{\partial I^a} \ ,
\label{formc}
\ee
whose integrability conditions can be expressed as
\be
\{X^a, Y_a\}_{\Phi E^b}  = 0\ , \quad \{X^a, Y_a\}_{\Phi I^b} = 0\ ,\nonumber
\ee
\be
 \{X^a, Y_a\}_{E^b I^c}  =  \frac{\delta_{bc}}{f} \ , 
\label{intc}
\ee
where 
\be
\{Z^{\bar A}, Z ^{\bar B}\}_{Z^A Z^B} = 
\frac{\partial Z^{\bar A} }{\partial Z^A} \frac{\partial Z^{\bar B} }{\partial Z^B} -
\frac{\partial Z^{\bar A} }{\partial Z^B} \frac{\partial Z^{\bar B} }{\partial Z^A}
 \ .
\ee 
The above integrability conditions resemble the conditions for a transformation 
of generalized coordinates and momenta to be a canonical transformation in classical mechanics. This is due to the 
fact that any section of ${\cal T}$ with fixed  $\Phi$ (or $F$) corresponds to a symplectic manifold with a symplectic 
structure similar to that of the phase space in classical mechanics.

We now consider the assumption (\ref{cont2}). It is straightforward to write down explicitly the corresponding system of partial differential equations in which we can insert the integrability conditions (\ref{intc}). As a result, we obtain
\bea
(X^a, Y_a)_{\Phi\Phi} = 
(X^a, Y_a)_{\Phi E^b} = 0 \ , \nonumber\\
(X^a, Y_a)_{\Phi I^b} =  
(X^a, Y_a)_{E^b E^c} = 0 , \nonumber \\ 
(X^a, Y_a)_{E^b I^c} = \frac{1}{2} h_{bc} - \frac{\delta_{bc}}{f} ,
\label{metc}
\eea
where  the round parenthesis represent the operator 
\be 
(Z^{\bar A}, Z ^{\bar B})_{Z^A Z^B} = \frac{\partial Z^{\bar A} }{\partial Z^A} \frac{\partial Z^{\bar B} }{\partial Z^B} \ .
\ee
 We thus 
have derived in Eqs.(\ref{formc}) and (\ref{metc}) the explicit differential equations  for the contactomorphism conditions (\ref{cont1}) and (\ref{cont2}) to be satisfied.

If we now insert the different values of the components $h_{ab}$ into the above differential equations, lengthy calculations show that they are not satisfied in general for any of the GTD-metrics, the integrability conditions being one of the main problems. Nevertheless, 
a detailed study of the analytic form of the integrability conditions show that they are satisfied if we allow the contact form to be subject to  a ``deformation'' of the form 
\be
\bar \Theta \rightarrow f_0dF - f_a Y_a d X^ a\ ,
\ee
where the non-vanishing  functions $f_0$ and $f_a$ can depend on all coordinates $F$, $X^a$ and $Y^a$. In other words, if we relax the contactomorphism condition (\ref{cont1}), and instead
assume that 
\be
\chi_{_{\cal T}}(d\Phi - I_a dE^a)  = f_0 dF - f_aY_a dX^a  \ , 
\label{cont1d}  
\ee
the integrability conditions are satisfied. It then follows that condition (\ref{cond1}) should be ``deformed" into   
\be
\bar G = (f_0 dF - f_a Y_a dX^a)^2 + \delta_{ab} dX^a dY ^b \ ,
\ee
for the integrability conditions to be satisfied.  Consequently, we demand that the line element for the phase space ${\cal T}$  
in the new coordinates $ Z ^ {\bar A}$ has exactly this form in accordance with the requirements of the integrability conditions 
in terms of the deformed contactomorphism.

Let us consider the corresponding equilibrium space ${\cal E}$ by means of the 
embedding map $\bar \varphi_{_X}:  {\cal E} \rightarrow {\cal T}$ which is defined by the condition 
$\bar \varphi_{_X}^*(dF - Y_a d X^a) = 0$. Then, the thermodynamic 
metric $\bar g= \bar \varphi_{_X}^*(\bar G)$ induced in ${\cal E}$ can be expressed as
\be
\bar g = (f_0-f_a)(f_0-f_b) \frac{\partial F}{\partial X^a} \frac{\partial F}{\partial X^b } dX^a dX ^b
+ \frac{\partial^2 F}{\partial X^a\partial X^b} dX^a dX ^b \ .
\ee 
We see that this metric is not Hessian because it  contains also first-order derivatives of the new coordinate $F(X^a)$. 
Moreover, it depends on $n+1$ arbitrary functions $f_0$ and $f_a$. Nevertheless, if we now demand that $F(X^a)$ 
reaches an extremal value at equilibrium, i.e., 
$\partial F/\partial X^a=0$ with $\frac{\partial^2 F}{\partial X^a\partial X^b}\neq 0$
at each point of the equilibrium manifold, then the metric $\bar g$ reduces to 
\be
\bar g =    \frac{\partial^2 F}{\partial X^a\partial X^b} dX^a dX ^b \ ,
\label{hessf}
\ee
i.e., it becomes Hessian. As discussed in Sec. \ref{sec:geomet}, to interpret the function $F(X^a)$ as a thermodynamic potential from a physical point of view further requirements must be  imposed, namely, that $F$ corresponds to the potential of the entire universe, whereas 
$X^a$ corresponds to the thermodynamic subsystem described by the metric (\ref{hessf}).

This result shows that for the metric of the equilibrium manifold to become Hessian, it is necessary not only that  
the new coordinate $F(X^a)$ be introduced as an element of a  diffeomorphism, but also that it can be interpreted as 
thermodynamic potential, according to the mathematical definition given in Sec. \ref{sec:geomet}. 
 Indeed, consider the infinitesimal fluctuations $dX^a$ of the potential $F(X^a)$ around an equilibrium state $X^a$. Then, taking into account that at equilibrium $F(X^a)$ is extremal,  we obtain up to the second approximation order  the Taylor expansion 
\be
F(X^a+dX^a) = F(X^a) + \frac{1}{2}   \frac{\partial^2 F}{\partial X^a\partial X^b} dX^a dX ^b \ .
\ee
We conclude that the components of the GTD-metrics in the equilibrium space can be interpreted as the second fluctuation moment of the new thermodynamic potential $F$, which is generated by a diffeomorphism of the phase space, satisfying the condition 
(\ref{cont1d}) of a deformed contactomorphism, $\chi_{_{\cal T}}(d\Phi - I_a dE^a)  = f_0 dF - f_aY_a dX^a$, 
 and acting  on the second part of the general GTD-metric as 
\be
\chi_{_{\cal T}}(h_{ab}dE^a dI^b)  =   \delta_{ab} dX^a dX^b\ .
\label{cond3}
\ee
Consider now the last condition. One can see that $\bar h = \delta_{ab} dX^a dX^b$ represents a well-defined line element in a $2n-$dimensional manifold ${\cal C}$, with coordinates $\{E^a, I^a\}$, which is known as the control manifold. Since in this particular case
the metric $\delta_{ab}$ is constant, the control manifold is flat. This implies that the curvature of the metric 
$h = \chi_{_{\cal T}}^{-1}(\bar h)=h_{ab}dE^a dI^b $ must vanish as well. A direct computation of the GTD-metrics shows that this condition is
satisfied only by the metric $G^{^{III}}$. Indeed, in this case we have that only the diagonal components of $h_{ab}$ are non-vanishing and are given as $ h_{aa} = (E_aI_a)^{2k+1}$; consequently, the corresponding manifold is flat. 

Condition (\ref{cond3}) implies that in the case of the metric $G^{^{III}}$ the diffeomorphism is determined by the relations
\be
F=\Phi, \quad X^a = \frac{(E^a)^{2k+2}}{2k+2}\ , \quad  Y^a = \frac{(I^a)^{2k+2}}{2k+2}\ , 
\ee
whereas the deformed contactomorphism is given by 
\be
f_0 = 1\ , \quad f_a = (2k+2) ^{ -\frac{2k+1}{k+1} } (X^aY^a)^{-\frac{2k+1}{2k+2}} \ .
\ee  
We thus see that in the case of the metric $G^{^{III}}$, the new thermodynamic potential $F(X^a)$ coincides with the original potential 
$\Phi$; however, its explicit functional dependence on the coordinates is different. 

It is easy to see that in the case of the GTD-metrics $G^{^I}$ and $G^{^{II}}$, the second part  of the metric, $h_{ab}dE^adI^b$, does not correspond to a flat control manifold. It is therefore necessary to apply a different approach.

\subsection{Hessian manifolds}
\label{sec:hess}

According to the results presented in the last subsection, the GTD-metrics $G^{^I}$ and $G^{^{II}}$, which are invariant with respect to total Legendre transformations, cannot induce a Hessian metric in ${\cal E}$, independently of coordinate system chosen in the phase manifold ${\cal T}$. We therefore consider the general form of the induced metric in terms of the extensive variables $E^a$. From  
Eq.(\ref{gupap}), we obtain 
\be
g^{^{I/II}} = \varphi^*(G^{^{I/II}})= \left( \xi_a^b E^a\frac{\partial\Phi}{\partial E^b}\right) 
\left( \chi_a^c\frac{\partial^2\Phi}{\partial E^b \partial E^c } dE^adE^b\right) \ , 
\label{gdown1}
\ee
with
\be
\xi_a^b = \xi_{ac}\delta^{cb}\ ,\qquad \chi_a^b = \chi_{ac}\delta^{cb}\ .
\ee
If, in addition, we consider the Euler identity for homogeneous or quasi-homogeneous  systems, the conformal factor becomes proportional to the thermodynamic potential itself \cite{qqs19} so that  the 
metrics (\ref{gdown1}) can be expressed, up to a multiplicative constant,  as 
\be
g^{^{I/II}} = \varphi^*(G^{^{I/II}})= \Phi \left( \chi_a^c\frac{\partial^2\Phi}{\partial E^b \partial E^c } dE^adE^b\right) \ .
\ee

The question we need to ask now is whether there exists a coordinate transformation $E^a \rightarrow X^a=X^a(E^b)$ such that the above metric 
becomes Hessian, i.e., it transforms into
\be
g^{^{I/II}} = \frac{\partial^2 F^{^{I/II}} }{\partial X^a \partial X^b } dX^a dX^b\ ,
\label{hessI}
\ee
where $F^{^{I/II}}=F^{^{I/II}}(X^a)$ is the Hessian potential. To  answer this question, it is necessary to invoke the theory of affine connections and dual connections on the equilibrium space ${\cal E}$. It turns out that a simple answer is not possible because it depends on the dimension of ${\cal E}$. In fact, the following theorems must be used to investigate this question \cite{aa13}.

{\bf Theorem 1.} All analytic 2-dimensional metrics are Hessian.

If the system under consideration is characterized by two thermodynamic degrees of freedom ($n=2$), and the fundamental equation 
$\Phi=\Phi(E^1,E^2)$ is an analytic function, the condition of the theorem is satisfied and, consequently, there exists a new potential 
$F^{^{I/II}}=F^{^{I/II}}(X^1,X^2)$, with $X^1=X^1(E^1,E^2)$ and $X^2=X^2(E^1,E^2)$ in which the metric becomes Hessian (\ref{hessI}).
Many important thermodynamic systems are included in this class;  for instance, the ideal gas and the van der Waals gas \cite{qsv15} with 
a fixed number of particles, the Reissner-Nordstr\"om and the Kerr black hole \cite{aqs08}, all the perfect fluids that are used in relativitic cosmology \cite{abcq12}, etc.

{\bf Theorem 2.} In 3 dimensions, all possible Riemann curvature tensors occur as the curvature tensor of a Hessian metric.

In the case of a system with three thermodynamic degrees of freedom ($n=3$), this theorems shows that the curvature of the corresponding thermodynamic metric is equivalent to the curvature of a Hessian metric. Nevertheless, it does not imply the existence of the underlying
Hessian potential. Consequently, in such a case it is necessary to consider the explicit form of the Hessian curvature and use the integrability 
conditions to determine the Hessian potential by using, for instance, the Cartan formalism \cite{quev92}.

{\bf Theorem 3.} In 4 and higher dimensions, for a given metric to be Hessian it is necessary that its Riemann curvature tensor satisfies the conditions
\be
\alpha(R_{ija}^{\ \ \ b} R_{klb}^{\ \ \ a}) = 0 \ ,
\ee 
\be
\alpha(R_{iajb} R_{k\ cd}^{\ b} R_l^{\ dac} - 2 R_{iajb} R_{kc\ d}^{\ a} R_l^{\ dbc}) = 0\ ,
\ee
where the operator $\alpha$ denotes antisymmetrization of the indices $i,j,k,l$. The above conditions are equivalent to demanding that the Pontryagin forms, associated with the Pontryagin characteristic class, vanish on a Hessian manifold \cite{aa13}
. This is due to the fact that Hessian manifolds possess a very particular set of affine connections which annihilates the Pontryagin characteristic class.

If the number of thermodynamic degrees of freedom of the system is $n\geq 4$, the fulfillment of the above conditions can be considered as an indication of the existence of a Hessian potential. 

The above theorems can be used to find the Hessian potential $F^{^{I/II}}$ for any given thermodynamic metric $g^{^{I/II}}$. 
Only in the case of $n=2$, it is possible to prove the existence of the Hessian potential in general. In the remaining cases, additional 
conditions must be satisfied which can be corroborated only if the fundamental equation $\Phi=\Phi(E^a)$ is given explicitly. 
This means that in principle there could be systems with $n\geq 3$ for which no Hessian potential exists. This should be tested for each system separately.

If, in addition, the Hessian potential is characterized by reaching an extremum at equilibrium, it can  be interpreted also as a thermodynamic potential and, consequently, the components of the GTD-metric for the equilibrium space can be interpreted as the second
fluctuation moment of  the Hessian potential. 

\section{Classification of thermodynamic potentials}
\label{sec:pot}

For investigating the physical significance of the GTD-metrics, it was necessary to introduce new thermodynamic potentials by using coordinate transformations. To distinguish the different types of potentials that can be used in thermodynamics and in the geometric representations of thermodynamics, we propose to classify them as fundamental, Legendre and diffeomorphic potentials, according to the following definitions. 

{\it Fundamental potentials:} These are potentials which are given as fundamental equations, i.e, as functions that depend on extensive variables only $\Phi=\Phi(E^a)$. This means that there exist only two fundamental potentials which can be identified as the entropy $S$ 
and the internal energy $U$. In classical thermodynamics, the use of the fundamental potentials lead to the well-known entropy and energy 
representations, respectively \cite{callen}.  In thermodynamic geometry, the fundamental potentials entropy and internal energy are used 
as Hessian potentials to define the Ruppeiner and Weinhold metrics, respectively.

{\it Legendre potentials:} These are potentials $\tilde \Phi$ 
 that can be obtained from the fundamental potentials $\Phi=\Phi(E^a)$  by means of 
Legendre transformations defined as
\be
 \Phi = \tilde \Phi - \tilde E ^k \tilde I_k  \ ,\quad
 E^i = - \tilde I ^ {i} \ , \ \
E^j = \tilde E ^j,\quad
 I^{i} = \tilde E ^ i , \ \
 I^j = \tilde I ^j \ ,
 \label{leg1}
\ee 
where $i\cup j$ is any disjoint decomposition of the set of
indices $\{1,...,n\}$, $k,l= 1,...,i$. Legendre potentials $\tilde \Phi$ 
are characterized by depending on at least one intensive variable. The limiting case in which it depends on only 
intensive variables, $\tilde\Phi =\tilde\Phi (I^a)$ corresponds to a total Legendre transformation for which $i=\{1,2,...,n\}$.
In general, for a system of $n$ thermodynamic degrees of freedom one can introduce up to $n^2-1$ different Legendre potentials.
In GTD, the main idea is that one can use any fundamental or Legendre potential to describe the geometric properties of the 
equilibrium manifold.

{\it Diffeomorphic potentials:} These are potentials that can be obtained by applying coordinate transformations 
on the GTD-metrics under the condition that the resulting metric for the equilibrium manifold is Hessian. There are two ways to 
generate diffeomorphic  potentials  by means of coordinate transformations either at level of the phase space, as explained in Sec. \ref{sec:def}, or 
at the level of the equilibrium space, as explained in Sec. \ref{sec:hess}. Diffeomorphic potentials are Hessian and simultaneously thermodynamic. 
They are important because the second moment of their fluctuation determines the components of the GTD-metrics. The diffeomorphic potentials appear 
as a consequence of the diffeomorphism invariance of the phase and equilibrium manifolds as defined in GTD. The number of possible diffeomorphic potentials is quite large. In principle, as shown in Sec. \ref{sec:def}, a Legendre potential can become a diffeomorphic
 potential after the application of a coordinate transformation on the phase manifold.

\section{Conclusions}
\label{sec:con}

In this work, we analyze the physical significance of the GTD-metrics from the point of view of fluctuation theory.
We have seen that imposing Legendre invariance and the condition that the equilibrium manifold of a non-interacting thermodynamic
system is flat, we end up with only three different metrics for the phase manifold, namely, $G^{^I}$ and $G^{^{II}}$, which are invariant with respect to total Legendre transformations, and $G^{^{III}}$, which is invariant with respect to arbitrary Legendre transformations.

We have shown that in the case of the metric $G^{^{III}}$, 
it is always possible to perform a coordinate transformation at the level of the phase manifold of GTD 
such that the metric induced in the equilibrium manifold becomes Hessian in terms of a new thermodynamic potential and new
 coordinates. Our analysis shows that the components of the GTD-metrics at the level of the equilibrium manifold
can be interpreted as the second fluctuation moment of the new thermodynamic potential with respect to the new  
coordinates. We found the explicit form of the new thermodynamic potential and the new coordinates for an arbitrary 
metric $G^{^{III}}$.

In the case of the  metrics $G^{^I}$ and $G^{^{II}}$, it is not possible to find the explicit form of the new thermodynamic potential in general.  Instead, for each thermodynamic system with a given fundamental equation the calculations must carried out separately at the level of the equilibrium manifold.  We use 
several theorems about the properties of Hessian manifolds to show that for systems with two degrees of freedom, it is always possible to find a Hessian potential. In the case of a higher number of degrees of freedom, the Riemann curvature of the equilibrium manifold must satisfy certain conditions in order for the  Hessian potential to exist. If, in addition, the Hessian potential possesses an extremum at equilibrium, it becomes the new thermodynamic potential which determines the physical significance of the GTD-metrics components.

The diffeomorphism invariance of the metrics used in GTD allows us to introduce a new type of thermodynamic potential which depends on coordinates not necessarily associated with the usual thermodynamic variables. Moreover, this result can be used to 
classify thermodynamic potentials into fundamental, Legendre and diffeomorphic potentials. Fundamental and Legendre potentials are used in classical thermodynamics which is invariant with respect to the change of potentials within this class. Diffeomorphic 
potentials are different because they are obtained by applying arbitrary coordinate transformations on the phase or equilibrium manifolds. The important point is that the components of the GTD-metrics correspond to the second fluctuation moment of diffeomorphic potentials. 

Whereas fundamental and Legendre potentials have a clear physical significance, diffeomorphic potentials have been introduced by using the mathematical properties of the phase and equilibrium manifolds. Their physical significance is therefore not yet completely clear. 
In the case of the metric $G^{^{III}}$, however, the diffeomorphic 
potential is just a fundamental or a Legendre potential written in terms of new coordinates; its physical meaning is therefore closely related to that of the known potentials of classical thermodynamics. In the case of 
the metrics $G^{^I}$ and $G^{^{II}}$, it is necessary to perform the explicit calculation of the diffeomorphic potential for each particular system 
in order to investigate its physical meaning.

\section*{Acknowledgments}

We would like to thank Prof. V. M. Romero-Roch\'\i n for very valuable comments and suggestions.
This work was partially supported
by UNAM-DGAPA-PAPIIT, Grant No. 111617.

\end{document}